\documentclass[acmsmall,nonacm=true,screen=true]{acmart}

\AtBeginDocument{%
  \providecommand\BibTeX{{%
    \normalfont B\kern-0.5em{\scshape i\kern-0.25em b}\kern-0.8em\TeX}}}

\setcopyright{acmcopyright}
\copyrightyear{202x}
\acmYear{202x}
\acmDOI{10.1145/123123123.121212}

\acmConference[arXiv]{arXiv '2x}{date}{arXiv}
\acmBooktitle{arXiv '2x, date, arXiv}
\acmPrice{00.00}
\acmISBN{978-1-4503-XXXX-X/2x/xx}

\usepackage{subfig}
\usepackage{tikz}
\usetikzlibrary{positioning, calc, shapes.geometric, shapes.symbols, arrows.meta, fit, backgrounds, intersections}
\usepackage{pgfplots}
\pgfplotsset{compat=1.16}
\usepackage{pgfplotstable}

\usepackage{xstring}

\usepackage{lmodern}
\begin{document}

\title{AI-in-the-Loop}
\subtitle{The impact of HMI in AI-based Application}

\author{Julius Schöning}
\email{j.schoening@hs-osnabrueck.de}
\affiliation{%
  \institution{Osnabrück University of Applied Sciences}
  \department{Faculty of Engineering and Computer Science}
  \city{Osnabrück}
  \country{Germany}
  \postcode{DE-49076}
}

\author{Clemens Westerkamp}
\email{c.westerkamp@hs-osnabrueck.de}

\affiliation{%
  \institution{Osnabrück University of Applied Sciences}
  \department{Faculty of Engineering and Computer Science}
  \city{Osnabrück}
  \country{Germany}
  \postcode{DE-49076}
}

\renewcommand{\shortauthors}{}
\renewcommand{\shorttitle}{}

\begin{abstract}
Artificial intelligence (AI) and human-machine interaction (HMI) are two keywords that usually do not fit embedded applications. Within the steps needed before applying AI to solve a specific task, HMI is usually missing during the AI architecture design and the training of an AI model. The human-in-the-loop concept is prevalent in all other steps of developing AI, from data analysis via data selection and cleaning to performance evaluation. During AI architecture design, HMI can immediately highlight unproductive layers of the architecture so that lightweight network architecture for embedded applications can be created easily. We show that by using this HMI, users can instantly distinguish which AI architecture should be trained and evaluated first since a high accuracy on the task could be expected. This approach reduces the resources needed for AI development by avoiding training and evaluating AI architectures with unproductive layers and leads to lightweight AI architectures. These resulting lightweight AI architectures will enable HMI while running the AI on an edge device. By enabling HMI during an AI uses inference, we will introduce the AI-in-the-loop concept that combines AI's and humans' strengths. In our AI-in-the-loop approach, the AI remains the working horse and primarily solves the task. If the AI is unsure whether its inference solves the task correctly, it asks the user to use an appropriate HMI. Consequently, AI will become available in many applications soon since HMI will make AI more reliable and explainable.
\end{abstract}








\maketitle

\section{Introduction}
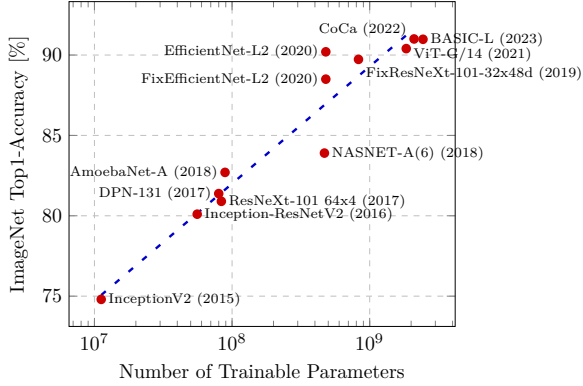
\begin{figure}[t!]
	\centering
	\scalebox{0.75}{
  \begin{tikzpicture}
\begin{axis}[xmode=log,
grid=major,
grid style={dashed,gray!50},
xlabel={Number of Trainable Parameters},
ylabel={ImageNet Top1-Accuracy [\%]},]
\addplot[
    red!80!black,
    thick,
    only marks,
    mark=*,
    visualization depends on=\thisrow{alignment} \as \alignment,
    nodes near coords,
    point meta=explicit symbolic,
    every node near coord/.style={anchor=\alignment,font=\scriptsize,black}
    ] table [
     meta=models,
     x=params,
     y=accuracy, col sep=comma
     ] {fig/ImageNetAccuracyVsParameters.csv};

   \addplot[draw=none] table [
        x=params,
        y=LinearRegressionValues,
        col sep=comma
        ] {fig/ImageNetAccuracyVsParameters.csv}
     coordinate [pos=1.0] (A)
     coordinate [pos=0.0] (B);

\draw[blue!80!black,loosely dashed, very thick] (A) -- (B);
\end{axis}
\end{tikzpicture}
  }
	\caption{A trend is visible when looking at selected state-of-the-art CNN architectures from recent years: A linear improvement in predictive performance coincides with an exponential increase in trainable parameters and, thus, computational complexity.}
	\label{fig:sota}\label{fig:deeper}
\end{figure}

Going deeper and deeper is a trend that can be recognized in AI architectures, e.g., for object detection tasks. As illustrated in Fig.~\ref{fig:deeper}, the number of parameters in convolutional neural networks (CNN) increased exponentially from InceptionV2~\cite{Ioffe2015} published in 2015 to BASIC-L~\cite{Chen2023} in 2023 by more than 115\%. This ongoing trend of deep artificial neural networks (ANN) leads to the need for more hardware resources, even for just the inference of an ANN. In addition, deeper ANN architectures also increase the hardware resources and the amount of data for training. Considering embedded AI-based applications, the question arises: "should one go deeper"~\cite{Richter2021} by developing AI architecture? To avoid deeper and deeper architecture, leading to the need for more computation on embedded hardware and more extensive data sets, this paper introduces intuitive human-machine interactions (HMI) and user interfaces (UI) for AI-based applications.

To identify fields of HMI in AI-based applications systematically, Section~\ref{sec:pipeline} gives a recap on the pipeline of applying AI for discussion in which steps intuitive HMI is standard and in which steps the potential of HMI and UI is still unleashed. By focusing on one step, where HMI is not yet common, Section~\ref{sec:anndesign} introduces and evaluates a UI focusing on the identification of unproductive CNN layers, ensuring lightweight ANN design. Based on lightweight ANN, a short excursus on AI for a typical task from embedded systems, the closed-loop control, is given in Section~\ref{sec:CLCS}. Making AI available for safety and security-critical applications, Section~\ref{sec:AIInTheLoop} introduces the AI-in-the-loop concept, where an interactive HMI will make embedded AI feasible within the upcoming years. Concluding this paper, Section~\ref{sec:conclusion}, the need for minimal HMI for applicable embedded AI-based applications is discussed, and further research directions are introduced.

\section{Pipeline of Applying AI}\label{sec:pipeline}
\begin{figure*}[t!]
  \centering
  \resizebox{0.95\linewidth}{!}
    {
  \input{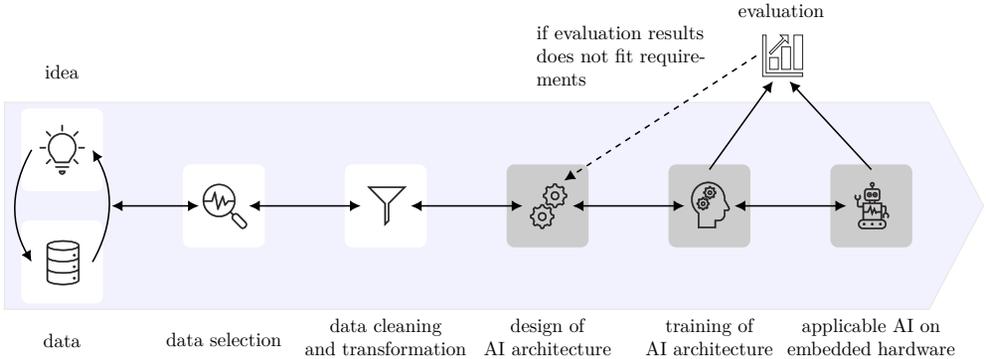}}
\caption{The pipeline of applying AI, interactive HMI is common in steps with white background; steps with a gray background are currently not supported by interactive HMI.}
\label{fig:pipeline}
\end{figure*}

In theory, the development of AI is a straightforward process composed of six steps. As illustrated in Fig.~\ref{fig:pipeline}, applying AI starts with data and an application idea. If the idea or the data came first is a chicken-and-egg problem. Usually, the ideas redefine the data collection and vice versa; the data trigger the ideas.

As soon as the application idea and the data lake are settled, selecting the data needed to solve the aimed application is the second step. In the first application iteration, by selecting the data for a noval application task, one should always ask, is a human capable of performing the tasks with the selected data? In our experience, asking this question will increase the success of the AI-based application in the early development phase. In later phases, especially in the context of embedded hardware, the data selection could be bolder so that the task might be impossible for a human being on the selection.

Based on the data selection, the next step in the pipeline is data cleaning and transformation. Due to AI making inferences based on statistical and mathematical procedures, balancing the dataset is necessary for avoiding biases and enforcing the generalization of the tasks by the ANN. Especially by processing image data of people, data cleaning for avoiding biases~\cite{Yang2020,Wang2019} in gender, ethnicity, and age is a crucial non-trial step. Next to cleaning, transforming the selected data in representation spaces suited for the AI architecture that should be applied is often mandatory. In the application case of predictive maintaining of, e.g., cooling fans~\cite{Westerkamp2020}, the data provided by the fans is usually a time series. Recurrent network architectures can be applied to time series classification tasks without a transformation. However, by transforming little sequences of the time series using, e.g., a Fourier transformation in a 2D representation, a CNN AI architecture can also be applied for predicting the malfunction. Thus, data transformation should always be considered for enabling different AI architectural designs

Determining which AI architecture fits the tasks and the data best is the entry point of this step. New kinds of AI architectures are published daily, so that an ANN zoo~\cite{Veen2019} is available to the developer. Before applying any architecture, one should ensure that the kind of AI architecture in mind is still state-of-the-art for covering the task. Next, to determine the kind of AI architecture, the number of layers and their parameters like kernel size, number of filters, and stride size will be defined, ending the architecture design.

The most resourceful and time-consuming step of the pipeline is the training of the AI architecture. The biggest subset of the selected, cleaned, and transformed data, normally 80\%, is used to train the parameter ANN. These trainable parameters, also known as weights, should generalize the tasks so that the inference of the ANN will work on unknown data. Another subset of data, normally 10\%, is used to evaluate the training performance during training. This evaluation is used to examine if and when overfitting occurs during the training. As soon as overfitting occurs, the ANN does not longer generalize the task; instead, the ANN starts to learn by rote.

The final step of the pipeline is applicable AI on embedded hardware. Thus, the AI-based application solves the desired tasks. Once the AI architecture reaches this step, the users come in contact with the AI-based application. Since users tend to explore the AIs' capacity, the AI-based application will often face adversarial attacks by the users~\cite{Huang2020,Fawaz2019}. These adversarial attacks cannot be prohibited; however, the operational design domain of the application should be clearly defined before the data selection step to provide data from all conditions occurring in the operational design domain.

So far, the pipeline is a straightforward process. Due to the number of trainable parameters of an ANN architecture, the task performance must be evaluated using unseen data. This subset of data, normally 10\%, is not used for any other step. It is exclusively for evaluation. If the evaluation with the unseen dataset does not fit the required performance, the design of the AI architecture needs to be refined, and the refined architecture must be trained again. This cycle of designing, training, and evaluating an AI architecture is very resource expensive.

Next to the evaluation step, interactive HMI and UI are only standard for the first three steps of the pipeline. For the last three steps, HMI and UI are not yet common.

%
%
%

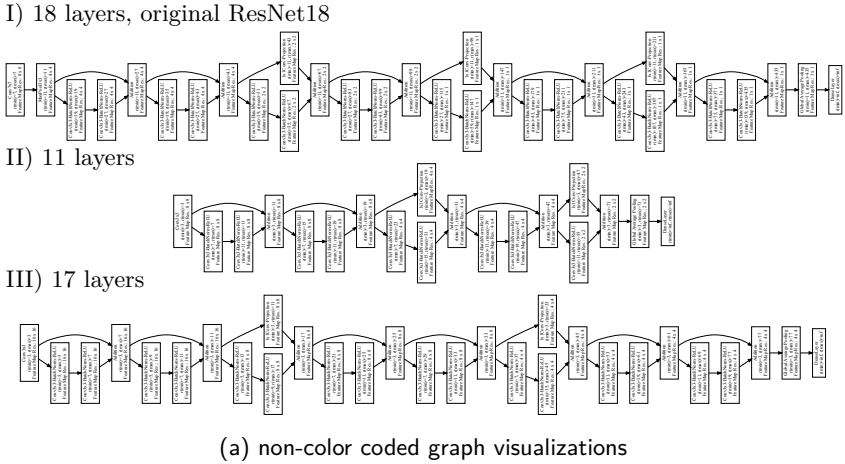
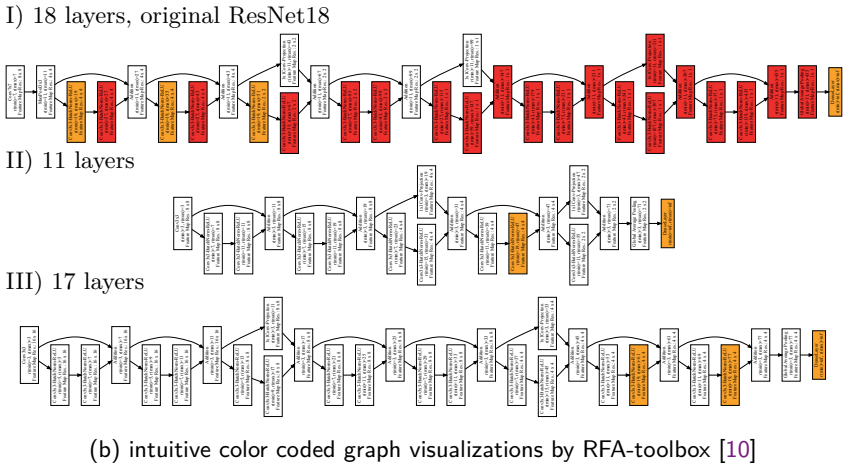
\begin{figure*}[t]
  \subfloat[non-color coded graph visualizations\label{fig:woUI}] {
  \begin{tikzpicture}
    \node(arch)[draw=none]
    {\includegraphics[width=0.8\textwidth]{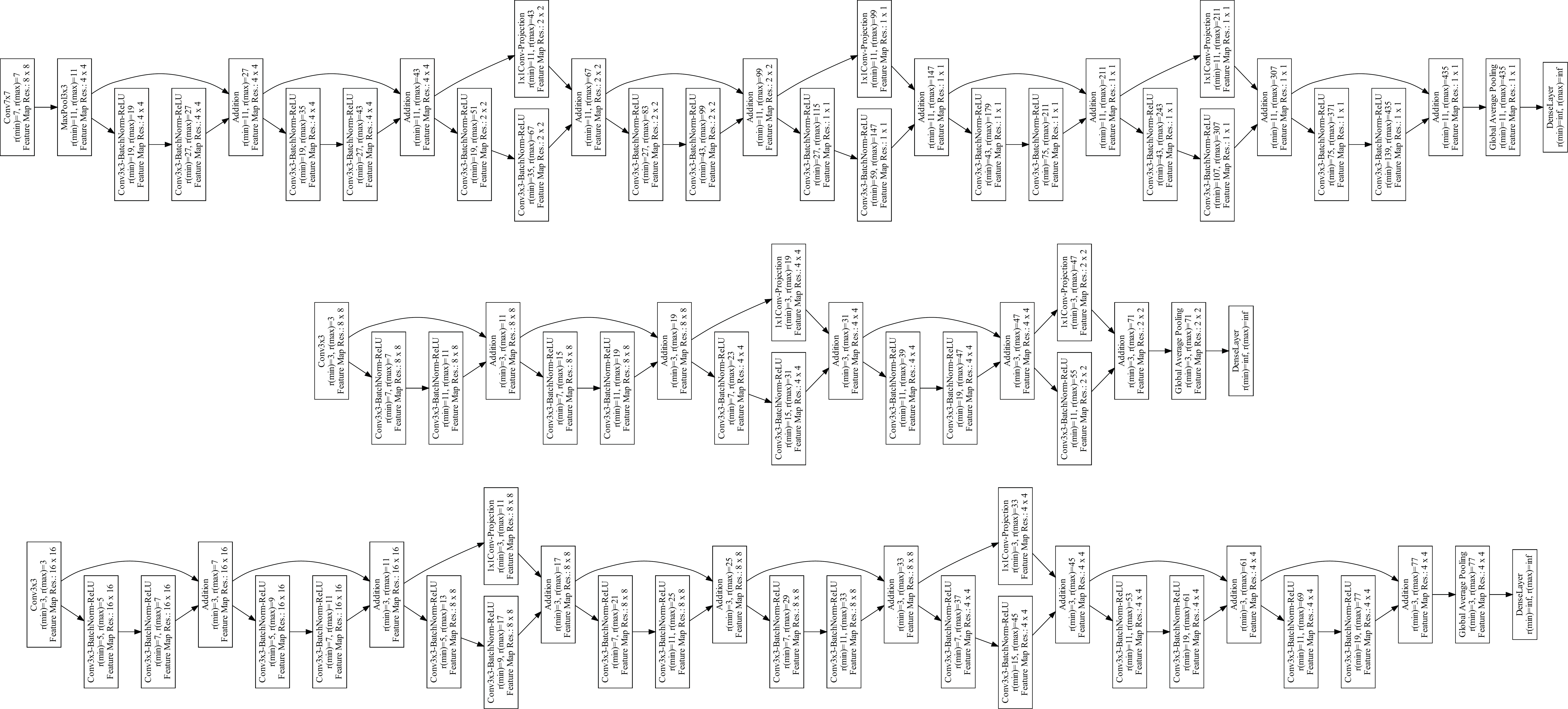}};
    \node(archTextI)[above left = 0.1cm and -0.0cm of arch, anchor = west] {\small I)  18 layers, original ResNet18};
    \node(archTextII)[above left = -1.8cm and -0.0cm of arch, anchor = west] {\small II) 11 layers};
    \node(archTextII)[above left = -3.4cm and -0.0cm of arch, anchor = west] {\small III) 17 layers};
  \end{tikzpicture}
  }

  \subfloat[intuitive color coded graph visualizations by RFA-toolbox~\cite{Richter2023}\label{fig:wUI}] {
  \begin{tikzpicture}
    \node(arch)[draw=none]
    {\includegraphics[width=0.8\textwidth]{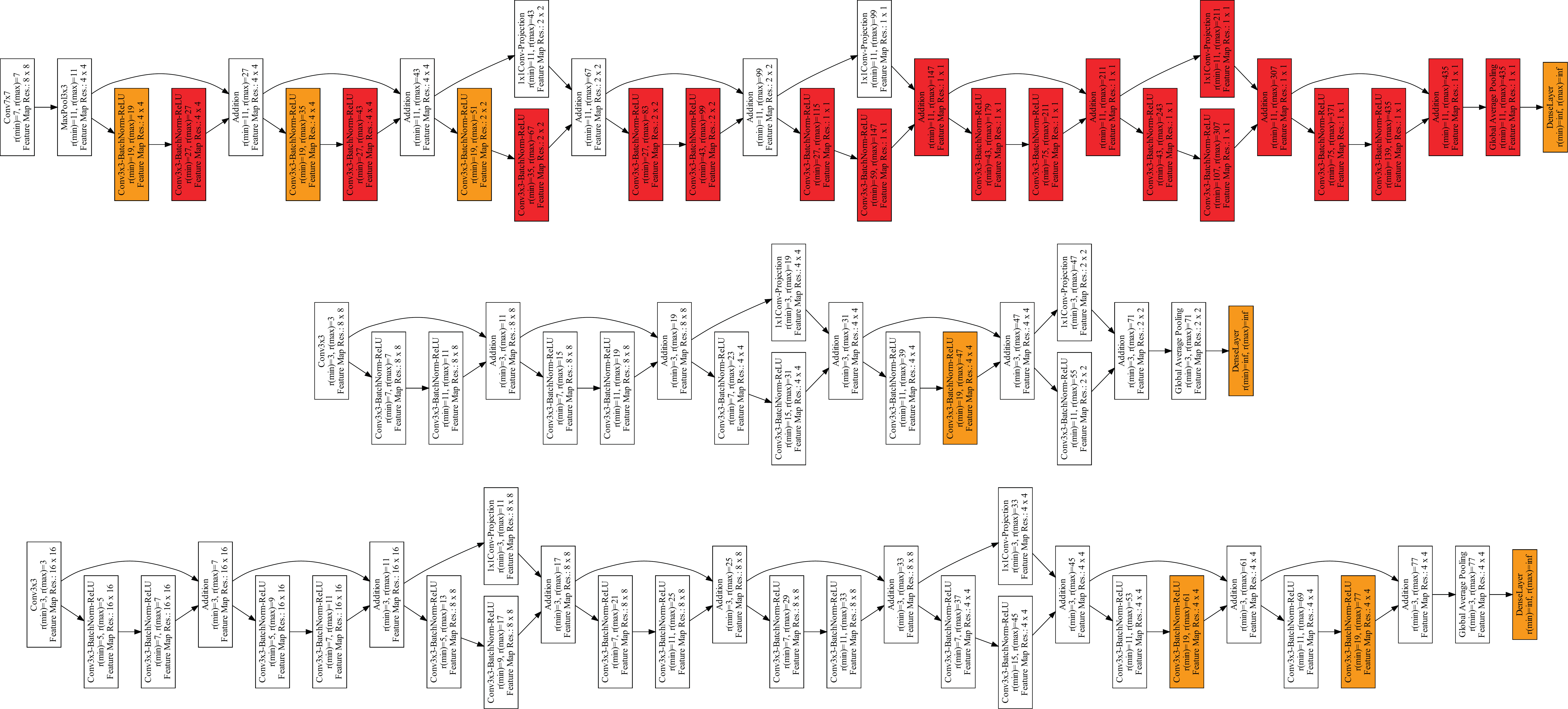}};
    \node(archTextI)[above left = 0.1cm and -0.0cm of arch, anchor = west] {\small I)  18 layers, original ResNet18};
    \node(archTextII)[above left = -1.8cm and -0.0cm of arch, anchor = west] {\small II) 11 layers};
    \node(archTextII)[above left = -3.4cm and -0.0cm of arch, anchor = west] {\small III) 17 layers};
  \end{tikzpicture}
  }
  	\caption{Three different AI architectures based on the ResNet18 architecture,  \protect\subref{fig:woUI} without highlighted unproductive layers within the architecture and \protect\subref{fig:wUI} with highlighted unproductive layers marked in red and border layer with limited productivity marked in orange.}\label{fig:task}
\end{figure*}

\begin{figure}[b]
    \centering
    \frame{\includegraphics[width=0.32\textwidth]{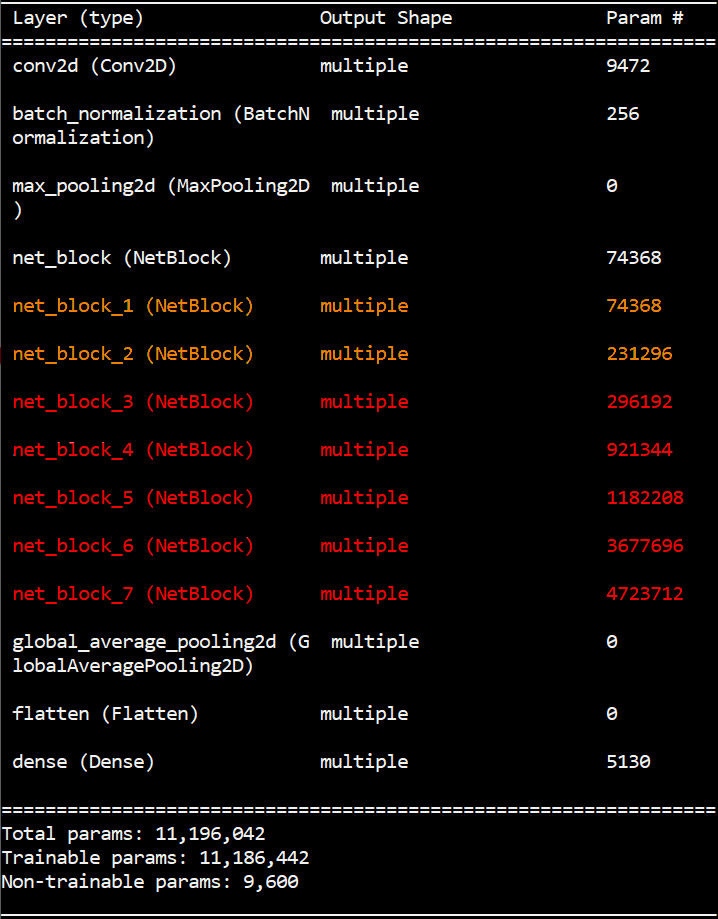}}
  	\caption{Command line visualization of unproductive layers by using color coding. Unproductive layers and groups of layers are marked in red, and border layers and groups with limited productivity are marked in orange.}\label{fig:newHMI}
\end{figure}

\section{Interactive Design of AI architecture} \label{sec:anndesign}
Enabling the user to instantly distinguish which AI architectures do not have unproductive layers, i.e., layers that do not influence the network performance without training the network, is valuable. Using the receptive field analysis (RFA), a UI can be provided to highlight these layers in CNN immediately.

\subsection{Receptive Field Analysis}
The area on the 2D input that influences the output of a convolutional layer is called a receptive field. Thus the multiple ways of the receptive fields within a CNN architecture can be computed. Using statistical methods, Luo et al.~\cite{Luo2016} determine the effective receptive field size in each layer, which can be considered to resemble a Gaussian distribution. Since Luo et al. computation method depends on the model's state, this method only works after the initialization of the architecture. Based on the topology of the architecture and the kernel and stride sizes of each layer, Koutini et al.~\cite{Koutini2019}, Wang et al.~\cite{Wang2020}, and Richter et al.~\cite{Richter2021} computational methods determine the receptive field size without initiation. The receptive field can generally be seen as a mathematical upper bound of theoretically extractable data features from the input. Thus, the receptive field is linked to the architecture's predictive quality, and computational efficiency~\cite{Richter2022}.

When a mismatch between the CNN architecture and the input data resolution occurs, the efficiency and predictive performance of the architecture is weak due to unproductive layers within the architecture. By avoiding unproductive layers, one can adjust the input data resolution to match the architecture to be between the minimum and maximum feasible input resolution; possible resolutions for common CNN architectures are in Tab.~\ref{tab:inversRFA} of the appendix. The other more envisaged way of avoiding unproductive layers is adjusting the CNN architecture. By removing unproductive layers, the number of trainable parameters decreases so that the training data tend to generalize better to the desired task during the training. As a side effect, the footprint of the architecture decreases, and the effort, both for human and computer during training and inference of these optimized architectures usually decreases, too.
\subsection{HMI for Identifying Unproductive Layers}
Users can create graph visualizations of CNN architectures with the easy-to-use RFA-Toolbox~\cite{Richter2023}, a python library. Using RFA, the toolbox marked all unproductive red and border layers with limited productivity orange, as illustrated in Fig.~\ref{fig:task}~\subref{fig:wUI}. The intuitive color coding UI of the graph visualizations allows the user to optimize the architectures without training. The users should optimize their CNN architecture until only productive, i.e., non-color coded, and border layer exists. Covering the needs of experienced AI developers even better, a command line visualization, as shown in Fig.~\ref{fig:newHMI} , is the next development step but has not yet been implemented.

\subsection{User Evaluation and Feedback}
An essential part of HMI and UI design is user evaluation. For this purpose, 16 computer science, electrical engineering, and mechatronics students evaluated the graph visualizations of the RFA-toolbox. On average, these subjects are in their 2.8th year of study, 24.7 years old, and have common knowledge of AI. All subjects were guided to eight questions in a structured interview. After gathering the demographical information on the age, the field of study, the year of study, and the experience in AI, AI-based image classification on the ImageNet data set~\cite{Russakovsky2015} was explained, ensuring the same knowledge level of the subjects. By showing the three different, non-color-coded ANN architectures of Fig.~\ref{fig:task}~\subref{fig:woUI}, the subjects were asked, "Which neural network solves the task most accurately?", "Which neural network solves the task most inaccurately?" and  "Which neural network has the most parameters?". After these first three questions, the intuitive color coding UI of the graph visualizations is introduced to the subjects. By showing the subjects the color-coded graph visualizations of the architectures, cf. Fig.~\ref{fig:task}~\subref{fig:wUI} following questions were asked "Which neural network solves the task most accurately?", "Which neural network solves the task most inaccurately?" and "Which neural network would you try first?". At the end of the interview, on a blank slide, the subjects were asked, "How helpful would you find such a software tool?" and "Do you think such a UI already exists?".

\begin{figure*}[t]
    \centering
  \subfloat[Which neural network solves the task most accurately?\label{fig:acc}] {
  \pgfplotstableread[col sep=comma]{
net,without,with
I),-11,4
II),-4,4
III),-1,8
}\loadedtable

\begin{tikzpicture}
\begin{axis}[
    width =0.8\textwidth,
    height = 1.5cm,
    name=popaxis,
    scale only axis,
    xbar,
    xmin = -16,
    xmax = 16,
    y dir = reverse,
    nodes near coords,
    bar width=8pt,
    bar shift=8pt,
    nodes near coords = {\pgfmathprintnumber[fixed,precision=0]\pgfplotspointmeta},
    every node near coord/.append style={font=\scriptsize, color=black},
    xtick={-15,-10,...,15},
    xticklabels= {15,10,5,0,5,10,15},
    symbolic y coords={I),II),III)},
    ytick={I),II),III)},
    y tick label style={yshift=0.3cm},
    axis x line=left,
    enlarge x limits = {value=0.15,upper},
    y axis line style={draw=none},
    x axis line style={-},
    y tick style={draw=none},
    clip=false
            ]
        \addplot [fill = gray!20] table[y=net,x=without] \loadedtable;
        \addplot [fill = gray!80] table[y=net,x =with] \loadedtable;
    \end{axis}
\end{tikzpicture}
  }
  \hfill
  \subfloat[Which neural network solves the task most inaccurately?\label{fig:inacc}] {
  \pgfplotstableread[col sep=comma]{
net,without,with
I),-5,12
II),-11,3
III),-0.001,1
}\loadedtable

\begin{tikzpicture}
\begin{axis}[
    width =0.8\textwidth,
    height = 1.5cm,
    name=popaxis,
    scale only axis,
    xbar,
    xmin = -16,
    xmax = 16,
    y dir = reverse,
    nodes near coords,
    bar width=8pt,
    bar shift=8pt,
    nodes near coords = {\pgfmathprintnumber[fixed,precision=0]\pgfplotspointmeta},
    every node near coord/.append style={font=\scriptsize, color=black},
    xtick={-15,-10,...,15},
    xticklabels= {15,10,5,0,5,10,15},
    symbolic y coords={I),II),III)},
    ytick={I),II),III)},
    y tick label style={yshift=0.3cm},
    axis x line=left,
    enlarge x limits = {value=0.15,upper},
    y axis line style={draw=none},
    x axis line style={-},
    y tick style={draw=none},
    clip=false
            ]
        \addplot [fill = gray!20] table[y=net,x=without] \loadedtable;
        \addplot [fill = gray!80] table[y=net,x =with] \loadedtable;
    \end{axis}
\end{tikzpicture}
  }
  \caption{Questions on the performance of the CNN architecture. In light gray, on the left side, the answers without the intuitive UI, and on the right side, the answers with the UI. Architecture III has the best accuracy on the task, and architecture I the worst.}\label{fig:pressionPlot}
\end{figure*}
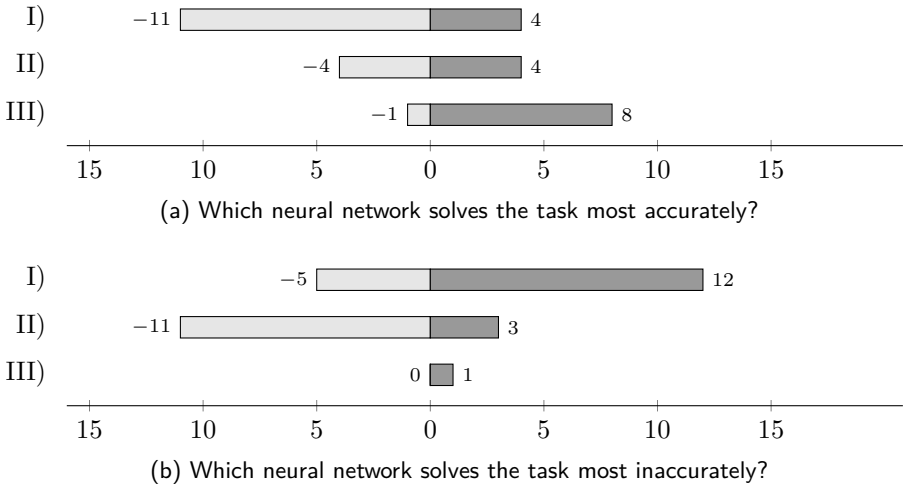

As depicted in Fig.~\ref{fig:pressionPlot}~\subref{fig:acc}, without the UI, 68.8\% of the subjects assumed that the 18 layer ResNet18 architecture solved the task most accurately. The general misbelief might support this conclusion that deeper ANN solves tasks always better, as the trend in state-of-the-art CNN architecture make belief. After introducing the UI 50\% of the subjects selected the 17 layer network---the correct answers which solved the image classification task most accurately. The cross-validation of this question, cf. Fig.~\ref{fig:pressionPlot}~\subref{fig:inacc}, also supports this results.

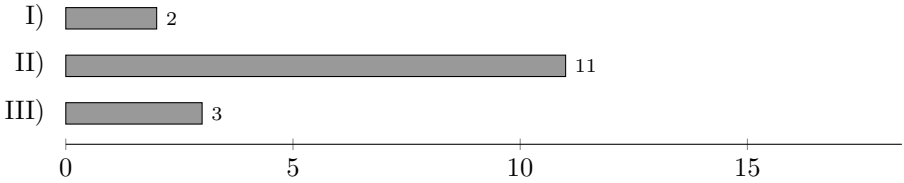
\begin{figure}[h]
  \centering
  \pgfplotstableread[col sep=comma]{
net,with
I),2
II),11
III),3
}\loadedtable

\begin{tikzpicture}
\begin{axis}[
    width =0.8\textwidth,
    height = 1.5cm,
    name=popaxis,
    scale only axis,
    xbar,
    xmin = 0,
    xmax = 16,
    y dir = reverse,
    nodes near coords,
    bar width=8pt,
    bar shift=8pt,
    nodes near coords = {\pgfmathprintnumber[fixed,precision=0]\pgfplotspointmeta},
    every node near coord/.append style={font=\scriptsize, color=black},
    xtick={0,5,...,15},
    xticklabels= {0,5,10,15},
    symbolic y coords={I),II),III)},
    ytick={I),II),III)},
    y tick label style={yshift=0.3cm},
    axis x line=left,
    enlarge x limits = {value=0.15,upper},
    y axis line style={draw=none},
    x axis line style={-},
    y tick style={draw=none},
    clip=false
            ]
        \addplot [fill = gray!80] table[y=net,x =with] \loadedtable;
    \end{axis}
\end{tikzpicture}
  \caption{Question: Which neural network would you try first? The answers collected with the UI of unproductive layers.}\label{fig:starting}
\end{figure}

The question "Which neural network has the most parameters?" was answered correctly by 87.5\% of the subjects. Only two subjects answered that the 11 layer architecture has the most parameters. Of the important answers on which architecture should be trained first, cf. Fig.~\ref{fig:starting},  68.8\% of the subjects answered the 11 layer architecture, which seems illogical since most subjects stated that the 17 layer architecture might be the most accurate one. After the interview, however, some subjects explained that the training of the smallest architecture is reasonable for evaluation due to the presence of computing power. The subjects argued that this smallest lightweight architecture might already fulfill the requirements of the desired tasks.

The question "How helpful would you find such a software tool?" all subjects answered that the RFA toolbox is quite helpful, and all subjects want to use this UI during the design of the AI architectures. However, only 50\% of the subjects think, that such a UI does not exist. The overall feedback on the HMI for designing AI architecture is very good.

\section{AI for Close Loop Applications}\label{sec:CLCS}
Lightweight ANN architectures will empower AI-based applications to reach new fields of embedded applications like closed-loop control systems (CLCS)~\cite{Schoening2022,Xie2020,Walczuch2022}, wearable cognitive enhancement technology~\cite{Robinson2006,Dubljevic2015,Schoening2023}, and interactive health devices~\cite{Abdi2015,Schoening2022a,Chang2019}. In addition, lightweight AI architectures are next to software design patterns~\cite{Schaarschmidt2020}, another aspect of energy optimization through software design. By introducing AI in CLCS, the functional safety of AI architectures has to be discussed. Until explainable AI is detailed delved, the safety dilemma of AI remains unsolved.
Thus AI-based controller needs to counter the risks. Without HMI, two possible designs can guarantee the functional safety of AI-based controllers. As illustrated in Fig.~\ref{fig:CLS_safety}~\subref{fig:CLS_safety_a}, one strategy is a fall-back path that disables the AI-based controller depending on, e.g., some threshold values. A second strategy is realized in Fig.~\ref{fig:CLS_safety}~\subref{fig:CLS_safety_b}, where a hand-crafted CLCS provides the control values, and the AI-based controller fine-tunes these values in predefined boundaries.

\begin{figure*}[t]
	\centering
	\subfloat[AI-based controller with a conventional fall-back path \label{fig:CLS_safety_a}] {
  \resizebox{0.8\linewidth}{!}
  {
  \begin{tikzpicture}
  \node(controller)[draw=red, ultra thick, rectangle, minimum width = 2.35cm, minimum height = 0.8cm] {\color{red}{\textbf{AI}}};

  \node(switch)[draw=red, ultra thick, right = 1.25cm of controller, rectangle, minimum width = 0.8cm, minimum height = 0.8cm] {};

  \node(AIswitch)[below = 0.3cm of switch, rectangle, minimum height = 0.8cm, inner sep=0pt]  {\color{red}\begin{tabular}{c}
  threshold \\ switch
  \end{tabular}
  };

  \node(process)[draw, right = 3.4cm of switch, rectangle, minimum width = 4.0cm, minimum height = 0.8cm] {\textbf{process}};

  \node(disturbances)[above = 0.3cm of process, rectangle, minimum height = 0.8cm] {\begin{tabular}{c}
  disturbances \\ \footnotesize $(D_{(t)})$
  \end{tabular}
  };

  \node(sensor)[draw, below right = 2.75cm and 3.35cm of controller, rectangle, minimum width = 4.0cm, minimum height = 0.8cm, anchor=center] {\textbf{sensor}};

  \node(response)[right = 0.7cm of process, rectangle, minimum height = 0.8cm] {\begin{tabular}{c}
   response \\ \footnotesize $(X_{(t)})$
  \end{tabular}
  };
  \node(sum)[left = 0.3cm of controller, circle, minimum height = 0.8cm] {
  };

  \node(reference)[left = 0.5cm of sum, rectangle, minimum height = 0.8cm] {\begin{tabular}{c}
   reference \\ \footnotesize $(W_{(t)})$
  \end{tabular}
  };

  \draw [-{Latex[length=2mm, width=2mm]}, ultra thick, draw=red]($(sensor.west)+(0,0.15)$)-|($(controller.south east)-(0.75,0)$);
  \node(controller2)[draw, below = 0.4cm of controller, thick, rectangle, minimum width = 2.35cm, minimum height = 0.8cm,fill=white] {{\textbf{controller}}};
  \node(parameter)[below = 0.3cm of controller2, rectangle, minimum height = 0.8cm, fill=white, inner sep=0pt]  {\begin{tabular}{c}
  static parameters \\ \footnotesize $(K_{P},K_{I},K_{D},\ldots)$
  \end{tabular}
  };

  \node(sum2)[draw, left = 0.3cm of controller2, circle, minimum height = 0.8cm] {
  };
  \draw [thick] (sum2.north east) -- (sum2.south west)
  (sum2.north west) -- (sum2.south east);

  \draw [-{Latex[length=2mm, width=2mm]}, thick, draw=black](switch.east)--(process.west) node [above,pos=0.5] (controlVariables) {control variables} node [below,pos=0.5] {\footnotesize $(X_{1(t)},X_{2(t)},X_{3(t)},\ldots)$};
  \draw [-{Latex[length=2mm, width=2mm]}, thick, draw=black](process.east)--($(response.west)+(0.3,0)$) node (rs) [circle, pos=0.3, fill=black, inner sep=0pt,minimum size=1mm] {};

  \draw [-{Latex[length=2mm, width=2mm]}, thick, draw=black]($(reference.east)+(-0.3,0)$)--(controller.west);
  \draw [-{Latex[length=2mm, width=2mm]}, thick, draw=black]($(disturbances.south)+(0,0.1)$)--(process.north);
  \draw [-{Latex[length=2mm, width=2mm]}, thick, draw=black]($(parameter.north)-(0,0.1)$)--(controller2.south);
  \draw [thick, draw=black](rs.center)|-(sensor.east);

  \draw [-{Latex[length=2mm, width=2mm]}, thick, draw=black]($(sensor.west)-(0,0.15)$)-|(sum2.south) node [right,pos=0.9] {\textbf{--}};

  \node (rsw) [left = 0.0cm of sum.center, circle, fill=black, inner sep=0pt,minimum size=1mm, anchor=center] {};
  \draw [-{Latex[length=2mm, width=2mm]}, ultra thick, draw=red](rsw.east)--(controller.west);
  \draw [-{Latex[length=2mm, width=2mm]},  thick, draw=black](rsw.center)--(sum2.north);
  \draw [-{Latex[length=2mm, width=2mm]}, thick, draw=black](sum2.east)--(controller2.west);

  \coordinate [left = 0.6cm of switch]  (swInput);

  \draw [-{Latex[length=2mm, width=2mm]}, ultra thick, draw=red](controller.east)-|($(swInput)+(0,0.15)$)|-($(switch.west)+(0,0.15)$);

  \draw [-{Latex[length=2mm, width=2mm]}, ultra thick, draw=red](controller2.east)-|($(swInput)-(0,0.15)$)|-($(switch.west)-(0,0.15)$);

  \coordinate [left = -0.2cm of switch]  (swIn);

  \node (swU) [above = 0.15cm of swIn.center, circle, fill=red, inner sep=0pt,minimum size=1mm, anchor=center] {};
  \draw [thick, draw=red]($(swInput)+(0,0.15)$)--(swU.center);

  \node (swD) [below = 0.15cm of swIn.center, circle, fill=red, inner sep=0pt,minimum size=1mm, anchor=center] {};
  \draw [thick, draw=red]($(swInput)-(0,0.15)$)--(swD.center);

  \coordinate [right = -0.2cm of switch]  (swOut);
  \node (swO) [below = 0.0cm of swOut.center, circle, fill=red, inner sep=0pt,minimum size=1mm, anchor=center] {};
  \draw [thick, draw=red](switch.east)--(swO.center);
  \draw [thick, draw=red](swU.center)--(swO.center);

  \node(AI)[above = 2cm of controller, rectangle, minimum width = 4.0cm, minimum height = 0.8cm] {};

  \node(recordedData)[draw=red, ultra thick, left = 0.5cm of AI, rectangle, minimum width = 3.0cm, minimum height = 0.8cm] {\color{red}{\begin{tabular}{c}\textbf{recorded process}\\ \textbf{data}
  \end{tabular}
  }};

  \draw [-{Latex[length=2mm, width=2mm]}, draw=red, ultra thick] (recordedData.east)-|($(controller.north)-(0.15,0)$);

  \draw [-{Latex[length=2mm, width=2mm]}, draw=red, ultra thick, dashed] ($(disturbances.west)+(0.4,0)$)-|($(controller.north)+(0.15,0)$)node [above, pos=0.25] {\color{red}{\textbf{prediction}}};

  \end{tikzpicture}
  }
  }

  \subfloat[AI-based controller with a limited impact on the control values \label{fig:CLS_safety_b}]  {
  \resizebox{0.8\linewidth}{!}
  {
  \begin{tikzpicture}
  \node(controller)[draw, rectangle, minimum width = 2.35cm, minimum height = 0.8cm] {\textbf{controller}};

  \node(parameter)[above = 0.3cm of controller, rectangle, minimum height = 0.8cm] {\begin{tabular}{c}
  static parameters \\ \footnotesize $(K_{P},K_{I},K_{D},\ldots)$
  \end{tabular}
  };

  \node(sum2)[draw=red, ultra thick, right = 1.25cm of controller, circle, minimum height = 0.8cm] {
  };
  \draw [draw=red, ultra thick] (sum2.north east) -- (sum2.south west)
  (sum2.north west) -- (sum2.south east);

  \node(sumtext)[below = 0.0cm of sum2, rectangle, minimum height = 0.8cm] {\color{red}\begin{tabular}{c}
  fine tune  \\ control variable \\  in predefined \\  boundaries
  \end{tabular}
  };

  \node(process)[draw, right = 3.4cm of sum2, rectangle, minimum width = 4.0cm, minimum height = 0.8cm] {\textbf{process}};

  \node(disturbances)[above = 0.3cm of process, rectangle, minimum height = 0.8cm] {\begin{tabular}{c}
  disturbances \\ \footnotesize $(D_{(t)})$
  \end{tabular}
  };

  \node(sensor)[draw, below right = 2.75cm and 3.35cm of controller, rectangle, minimum width = 4.0cm, minimum height = 0.8cm, anchor=center] {\textbf{sensor}};

  \node(response)[right = 0.7cm of process, rectangle, minimum height = 0.8cm] {\begin{tabular}{c}
   response \\ \footnotesize $(X_{(t)})$
  \end{tabular}
  };
  \node(sum)[draw, left = 0.3cm of controller, circle, minimum height = 0.8cm] {
  };
  \draw [thick] (sum.north east) -- (sum.south west)
  (sum.north west) -- (sum.south east);

  \node(reference)[left = 0.5cm of sum, rectangle, minimum height = 0.8cm] {\begin{tabular}{c}
   reference \\ \footnotesize $(W_{(t)})$
  \end{tabular}
  };

  \node(controller2)[draw=red, above right = 0.4cm and 0.1cm of controller, ultra thick, rectangle, minimum width = 2.35cm, minimum height = 0.8cm,fill=white] {\color{red}{\textbf{AI}}};

  \draw [-{Latex[length=2mm, width=2mm]}, thick, draw=black](sum2.east)--(process.west) node [above,pos=0.5] (controlVariables) {control variables} node [below,pos=0.5] {\footnotesize $(X_{1(t)},X_{2(t)},X_{3(t)},\ldots)$};
  \draw [-{Latex[length=2mm, width=2mm]}, thick, draw=black](process.east)--($(response.west)+(0.3,0)$) node (rs) [circle, pos=0.3, fill=black, inner sep=0pt,minimum size=1mm] {};

  \draw [-{Latex[length=2mm, width=2mm]}, thick, draw=black]($(reference.east)+(-0.3,0)$)--(sum.west);
  \draw [-{Latex[length=2mm, width=2mm]}, thick, draw=black]($(disturbances.south)+(0,0.1)$)--(process.north);
  \draw [-{Latex[length=2mm, width=2mm]}, thick, draw=black]($(parameter.south)+(0,0.1)$)--(controller.north);
  \draw [thick, draw=black](rs.center)|-(sensor.east);

  \draw [-{Latex[length=2mm, width=2mm]}, thick, draw=black]($(sensor.west)-(0,0.15)$)-|(sum.south) node [right,pos=0.95] {\textbf{--}};
  \draw [-{Latex[length=2mm, width=2mm]}, ultra thick, draw=red]($(sensor.west)+(0,0.15)$)-|($(controller2.south)-(0.8,0)$);

  \node (rsw) [right = 0.2cm of controller, circle, fill=red, inner sep=0pt,minimum size=1.5mm, anchor=center] {};

  \draw [-{Latex[length=2mm, width=2mm]}, ultra thick, draw=red] let
  \p1=($(rsw.center)$), \p2=($(controller2.south)$) in (rsw.center)--(\x1,\y2);

  \draw [-{Latex[length=2mm, width=2mm]}, ultra thick, draw=red] let
  \p1=($(sum2.north)$), \p2=($(controller2.south)$) in (\x1,\y2)--(sum2.north)node [right,pos=0.5] {\color{red}\textbf{+}};

  \draw [-{Latex[length=2mm, width=2mm]}, thick, draw=black](sum.east)--(controller.west);
  \draw [-{Latex[length=2mm, width=2mm]}, ultra thick, draw=red](controller.east)--(sum2.west) node [above,pos=0.85] {\color{red}\textbf{+}};;

  \node(AI)[above = 2cm of controller, rectangle, minimum width = 4.0cm, minimum height = 0.8cm] {};

  \node(recordedData)[draw=red, ultra thick, left = 0.5cm of AI, rectangle, minimum width = 3.0cm, minimum height = 0.8cm] {\color{red}{\begin{tabular}{c}\textbf{recorded process}\\ \textbf{data}
  \end{tabular}
  }};

  \draw [-{Latex[length=2mm, width=2mm]}, draw=red, ultra thick] (recordedData.east)-|($(controller2.north)$);

  \draw [-{Latex[length=2mm, width=2mm]}, draw=red, ultra thick, dashed] ($(disturbances.west)+(0.4,0)$)--($(controller2.east)$)node [above, pos=0.5] {\color{red}{\textbf{prediction}}};

  \end{tikzpicture}
  }
  }
\caption{Controlling safety critical processes by AI-based controller without the need of explainable AI~\cite{Schoening2022}.}
\label{fig:CLS_safety}
\end{figure*}

Anyhow, to improve functional safety, the number of trainable weights should be kept to a minimum by designing the ANN. HMI, like the RFA toolbox, will help the developers in doing so. Less trainable parameters reduce the complexity within the ANN and bring the applicability as well as the explainability of ANN architectures a step closer.

\section{AI-in-the-Loop} \label{sec:AIInTheLoop}
\begin{figure}[t!]
    \centering
    \scalebox{0.7}{
    \begin{tikzpicture}
  \node(controller)[draw, rectangle, minimum width = 2.5cm, minimum height = 2.0cm] {\begin{tabular}{c}\textbf{data lake /}\\ \textbf{data hub} \end{tabular}};

  \node(AI)[draw=red,, right = 1.7cm of controller, ultra thick, rectangle, minimum width = 2.5cm, minimum height = 2.0cm] {\color{red}{\textbf{AI}}};

  \node(process)[draw, right = 1.7cm of AI, rectangle, minimum width = 2.5cm, minimum height = 2.0cm] {\begin{tabular}{c}\textbf{mechatronical / }\\ \textbf{mechatronical / }\\ \textbf{bioelectrical systems} \end{tabular}};

  \node(disturbances)[above = 0.3cm of process, rectangle, minimum height = 2.0cm] {\begin{tabular}{c}
  disturbances \\ \footnotesize $(D_{(t)})$
  \end{tabular}
  };

  \node(sensor)[draw, below = 2cm of AI, rectangle, minimum width = 4.0cm, minimum height = 2.0cm, anchor=center] {\begin{tabular}{c}\textbf{sensors /}\\ \textbf{smart devices} \end{tabular}};

  \node(response)[right = 0.7cm of process, rectangle, minimum height = 2.0cm] {\begin{tabular}{c}
   response \\ \footnotesize $(X_{(t)})$
  \end{tabular}
  };
  \node(sum)[draw, left = 0.7cm of controller, circle, minimum height = 0.8cm] {
  };
  \draw [thick] (sum.north east) -- (sum.south west)
  (sum.north west) -- (sum.south east);

  \node(reference)[left = 0.5cm of sum, rectangle, minimum height = 2.0cm] {\begin{tabular}{c}
   reference \\ \footnotesize $(W_{(t)})$
  \end{tabular}
  };

  \coordinate(body) at ($(AI.north)+(0,5.0)$);
  \begin{scope}[shift={(body)}]
         \draw[fill=white,thick](-0.5,-0.2) -- (0.5,-0.2) -- (0.8,-0.5) -- (0.8,-1.6) -- (0.5,-1.6) -- (0.5,-1) -- (0.5,-3) -- (0,-3) -- (0,-1.8) -- (0,-3) -- (-0.5,-3) -- (-0.5,-1) -- (-0.5,-1.6)coordinate (h1) -- (-0.8,-1.6)--(-0.8,-0.5)coordinate[midway] (h2) -- cycle;
          \draw[fill=white,thick] (0,0.35) circle (0.4cm)coordinate(hHead);
          \node(user)[above = 0.5cm of hHead, rectangle] {\textbf{user / domain expert}};
  \end{scope}


   \draw [line width=1pt, double distance=3pt, arrows = {Latex[length=0pt 3 0]-Latex[length=0pt 3 0]}, draw=blue] ($(AI.north)+(0,2.0)$) -- (AI.north) node [sloped,pos=0.5] (text) {};
   \node(hmi)[left = 0.1cm of text, rectangle] {\begin{tabular}{c}\color{blue}{\textbf{minimal HMI}} \end{tabular}};
   \node(hmi2)[above right = 0.4cm and 0.3cm of text, rectangle] {\begin{tabular}{c}\footnotesize {bits of advice,} \\ \footnotesize {feedback, failsafe} \\ \footnotesize {optimizations, etc.} \end{tabular}};

  \draw [-{Latex[length=2mm, width=2mm]}, thick, draw=none]($(controller.north east)+(0,0)$)--($(AI.north west)+(0,0)$) node [above,pos=0.5] (controlVariables) {fused data};
  \draw [-{Latex[length=2mm, width=2mm]}, thick, draw=black]($(controller.north east)+(0,-0.3)$)--($(AI.north west)+(0,-0.3)$) node [above,pos=0.5] (controlVariables) {};
  \draw [-{Latex[length=2mm, width=2mm]}, thick, draw=black]($(controller.south east)+(0,0.3)$)--($(AI.south west)+(0,0.3)$) node [above,pos=0.5] (controlVariables) {\begin{tabular}{c} \textbf{$\bullet$}  \\ \textbf{$\bullet$} \\ \textbf{$\bullet$} \\ \end{tabular}};
  \draw [-{Latex[length=2mm, width=2mm]}, thick, draw=none]($(controller.south east)+(0,0.0)$)--($(AI.south west)+(0,0.0)$)  node [below,pos=0.5] (controlVariables) {\begin{tabular}{c} \footnotesize $(d_{1(t)},d_{2(t)},d_{3(t)},\ldots)$ \end{tabular}};

  \draw [-{Latex[length=2mm, width=2mm]}, thick, draw=none]($(AI.north east)+(0,0)$)--($(process.north west)+(0,0)$) node [above,pos=0.5] (controlVariables) {control variables};
  \draw [-{Latex[length=2mm, width=2mm]}, thick, draw=black]($(AI.north east)+(0,-0.3)$)--($(process.north west)+(0,-0.3)$) node [above,pos=0.5] (controlVariables) {};
  \draw [-{Latex[length=2mm, width=2mm]}, thick, draw=black]($(AI.south east)+(0,0.3)$)--($(process.south west)+(0,0.3)$) node [above,pos=0.5] (controlVariables) {\begin{tabular}{c} \textbf{$\bullet$}  \\ \textbf{$\bullet$} \\ \textbf{$\bullet$} \\ \end{tabular}};
  \draw [-{Latex[length=2mm, width=2mm]}, thick, draw=none]($(AI.south east)+(0,0.0)$)--($(process.south west)+(0,0.0)$)  node [below,pos=0.5] (controlVariables) {\begin{tabular}{c} \footnotesize $(X_{1(t)},X_{2(t)},X_{3(t)},\ldots)$ \end{tabular}};

  \draw [-{Latex[length=2mm, width=2mm]}, thick, draw=black](process.east)--($(response.west)+(0.3,0)$) node (rs) [circle, pos=0.3, fill=black, inner sep=0pt,minimum size=1mm] {};
  \draw [-{Latex[length=2mm, width=2mm]}, thick, draw=black](sum.east)--(controller.west);
  \draw [-{Latex[length=2mm, width=2mm]}, thick, draw=black]($(reference.east)+(-0.3,0)$)--(sum.west);
  \draw [-{Latex[length=2mm, width=2mm]}, thick, draw=black]($(disturbances.south)+(0,0.1)$)--(process.north);
  \draw [thick, draw=black](rs.center)|-(sensor.east);
  \draw [-{Latex[length=2mm, width=2mm]}, thick, draw=black](sensor.west)-|(sum.south) node [right,pos=0.8] {\textbf{--}};

  \end{tikzpicture}}
  	\caption{AI-in-the-loop concept enables the use of safe, secure, and reliable AI---the AI is the working horse and primarily solves the task and, with the support of minimal HMI, users' domain expertises are integrated if needed.}\label{fig:aiitl}
\end{figure}
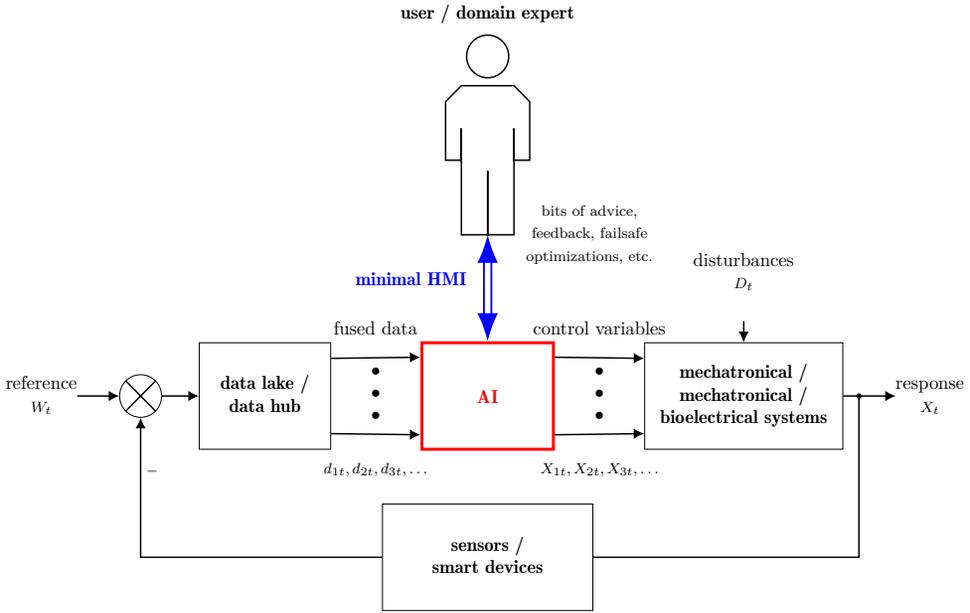

One often neglected solution is a minimal HMI for safeguarding the task execution in situations where the AI algorithm detects difficulties in inference. Merging the user-in-the-loop~\cite{Schoenen2014,Evers2014} concept with AI, AI will replace the user in the main loop. As illustrated in Fig.~\ref{fig:aiitl}, the AI-in-the-loop concept is based on the idea that the AI will do the majority of the workload. As the domain expert, the user will teach and support the AI in overcoming the safety, security, and task issues that the AI architectures cannot realize reliably. By introducing HMI in this step of the pipeline of applying AI, AI-based embedded applications can be available on the market sooner than fully automated AI-based applications.

The AI-in-the-loop combines AI's computational and human conceptual strengths by enabling HMI and UI during the AI's inferences. With its computational powers during inferences, the AI remains the working horse and primarily solves the task. If the AI is unsure whether it solves the task correctly or if the AI recognizes a safety or security-critical situation, it asks the user by using an appropriate HMI with an interactive UI. By enabling the development of intuitive HMI and UI, an AI architecture's trained high-dimensional task representation must be translated into a human-understandable problem description. First, UI methods, e.g., for visualizing as well as describing why the AI inferences the current output~\cite{Velden2022,Zeiler2014,Fong2017} already exist, but more research is needed to match the users' needs within the  AI-in-the-loop concept.

\begin{figure}[ht!]
  \centering
  \scalebox{0.7}{
  \begin{tikzpicture}
\renewcommand{\arraystretch}{0.7}
\begin{axis}[
axis lines=middle,
xlabel={\begin{tabular}{c}
  \\
  \footnotesize \textit{availability in} \\
  \footnotesize \textit{the market}
\end{tabular}},
ylabel={\begin{tabular}{c}
  \\
  \footnotesize \textit{users' workload} \\
  \footnotesize \textit{needed for} \\
  \footnotesize \textit{a complex task}
\end{tabular}},
xlabel style = {anchor=north},
ylabel style = {anchor=south},
ymax = 10.5,
xmax = 10.5,
xtick=\empty, ytick=\empty,
clip=false
           ]
\addplot [name path=A,domain=0:10,thick] {(2/3)^(x-5.7)};
\node(UIL) [circle,fill=blue,minimum width =0.2cm, minimum height =0.2cm, inner sep=0pt] at (1,6.72) {};
\node(UILt) [right = -.3cm  of UIL] {\begin{tabular}{c} \footnotesize user-in-the-loop \\ \footnotesize systems \end{tabular}};
\node(expert) [circle,fill=blue,minimum width =0.2cm, minimum height =0.2cm, inner sep=0pt] at (3.5,2.44) {};
\node(expertt) [left = -.3cm  of expert] {\begin{tabular}{c} \footnotesize AI-based \\ \footnotesize expert systems \end{tabular}};
\node(AIL) [circle,fill=blue,minimum width =0.2cm, minimum height =0.2cm, inner sep=0pt] at (5,1.33) {};
\node(AILt) [above right = -.3cm and -.4cm  of AIL] {\begin{tabular}{c} \footnotesize AI-in-the-loop \\ \footnotesize systems \end{tabular}};
\node(Au) [circle,fill=blue,minimum width =0.2cm, minimum height =0.2cm, inner sep=0pt] at (10,0.2) {};
\node(Aut) [above left= -0.1cm and -1.0cm  of Au] {\begin{tabular}{c} \footnotesize fully automated \\ \footnotesize AI-based system \end{tabular}};

\draw[dashed] (4.5,10) -- (4.5,0) node[below] {\footnotesize today};
\end{axis}
   \end{tikzpicture}
  }
  	\caption{Illustration of users' workload needed for performing a complex task over the availability of systems.}\label{fig:timeline}
\end{figure}
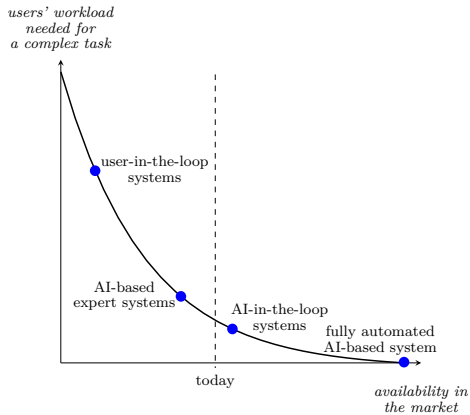

In combination with an interactive design of AI architectures and removing unproductive layers for ensuring lightweight ANN, the development of minimal HMI supporting AI-in-the-loop, will enable the use of safe and reliable AI in the near further. Fig.~\ref{fig:timeline} plots the users' workload needed for performing a complex task over the available of systems. Before reaching a full automatization, i.e. the users' workload do not longer exsist, it may be assumed that human AI cooperation in AI-in-the-loop-based systems will the upcoming logical development step. Until explainable AI is not detailed delved, AI-based applications have unknown blind spots.

\section{Conclusion}\label{sec:conclusion}
The pipeline of applying AI, as shown in Fig.~\ref{fig:pipeline}, reveals three steps where Interactive HMI is not common yet. As introduced and evaluated in this work, RFA and other methods improve the step of designing lightweight AI architecture by using HMI. Lightweight AI architectures are the enabler for explainable and interactive AI-based applications, following the  AI-in-the-loop concept. Instead of providing fully automated AI-based systems, one should focus on cooperative AI architecture, which uses both the AI's computational power and human conceptual thinking.

One blind spot for interactive HMI is still the step of training AI architecture; nowadays, developers visualize the accuracy and loss of an architecture undertraining. These visualizations are then, next to defined thresholds used for stopping the training. By using, e.g., virtual reality, developers might be capable of walking through the architecture and seeing how the backpropagations shape the trainable parameters. Understanding how each backpropagation shapes the trainable parameters might enable the developers to design better architectures, e.g., that generalize from only a few data points or help developers to optimize the hyperparameter so that, e.g., the local minimum can be avoided.

By concluding this work, one should consider embedded AI-based applications with minimal HMI as the next step toward a fully automated AI-based system. Going step by step and using the AI-in-the-loop concept will make applications sooner available in the market. With the experiences gained, having these applications in the market, safe, secure, and reliable AI will be feasible.




\bibliographystyle{myIEEEtran}
\bibliography{ref}

\pagebreak
\section*{Appendix A}

\begin{table*}[h]
  \caption{Minimum and maximum feasible input resolution $(I_{min}, I_{max})$ on \linebreak \textit{tensorflow} and \textit{torch} built-in CNN implementation. \textsuperscript{$\dagger$} resulting values of \textit{tensorflow} and \textit{torch} implementation differs \cite{Richter2022}. \label{tab:inversRFA}}

\resizebox{\textwidth}{!}{
\renewcommand{\arraystretch}{1.25}
\begin{tabular}{
>{\raggedright\arraybackslash}p{0.20\linewidth}|
>{\centering\arraybackslash}p{0.15\linewidth}|
>{\centering\arraybackslash}p{0.15\linewidth}|
>{\raggedright\arraybackslash}p{0.01\linewidth}
>{\raggedright\arraybackslash}p{0.27\linewidth}|
>{\centering\arraybackslash}p{0.15\linewidth}|
>{\centering\arraybackslash}p{0.15\linewidth}|
}
  \textbf{architecture}
  &
  $I_{min}$ &
  $I_{max}$ &
  &
  \textbf{architecture}
  &
  $I_{min}$ &
  $I_{max}$   \\ \cline{1-3} \cline{5-7} \noalign{\vskip\doublerulesep
         \vskip-\arrayrulewidth} \cline{1-3} \cline{5-7}
  \textbf{DenseNet121} &
  $103\times103$ &
  $2071\times2071$ &
  &
  \textbf{MobileNet} &
  \multicolumn{2}{c|}{$315\times315$} \\ \cline{1-3} \cline{5-7}

  \textbf{DenseNet161} &
  $103\times103$ &
  $2967\times2967$ &
  &
  \textbf{MobileNetV2} &
  $163\times163$ &
  $491\times491$ \\ \cline{1-3} \cline{5-7}

  \textbf{DenseNet169} &
  $103\times103$ &
  $3351\times3351$ &
  &
  \textbf{MobileNetV3 large} &
  $263\times263$ &
  $595\times595$ \\ \cline{1-3} \cline{5-7}

  \textbf{DenseNet201} &
  $103\times103$ &
  $3863\times3863$ &
  &
  \textbf{MobileNetV3 small} &
  $303\times303$ &
  $639\times639$ \\ \cline{1-3} \cline{5-7}

  \textbf{EfficientNetB0} &
  $299\times299$ &
  $851\times851$ &
  &
  \textbf{MnasNet} &
  $283\times283$ &
  $843\times843$ \\ \cline{1-3} \cline{5-7}

  \textbf{EfficientNetB1} &
  $299\times299$ &
  $1183\times1183$ &
  &
  \textbf{NASNetLarge} &
  $327\times327$ &
  $3303\times3303$ \\ \cline{1-3} \cline{5-7}

  \textbf{EfficientNetB2} &
  $299\times299$ &
  $1183\times1183$ &
  &
  \textbf{NASNetMobile} &
  $327\times327$ &
  $2407\times2407$ \\ \cline{1-3} \cline{5-7}

  \textbf{EfficientNetB3} &
  $299\times299$ &
  $1407\times1407$ &
  &
  \textbf{ResNet18} &
  $139\times139$ &
  $435\times435$ \\ \cline{1-3} \cline{5-7}

  \textbf{EfficientNetB4} &
  $299\times299$ &
  $1799\times1799$ &
  &
  \textbf{ResNet34} &
  $139\times139$ &
  $899\times899$ \\ \cline{1-3} \cline{5-7}

  \textbf{EfficientNetB5} &
  $299\times299$ &
  $2131\times2131$ &
  &
  \textbf{ResNet50\textsuperscript{$\dagger$} } &
  $96\times96$ &
  $427\times427$ \\ \cline{1-3} \cline{5-7}

  \textbf{EfficientNetB6} &
  $299\times299$ &
  $2523\times2523$ &
  &
  \textbf{ResNet101\textsuperscript{$\dagger$} } &
  $96\times96$ &
  $971\times971$ \\ \cline{1-3} \cline{5-7}

  \textbf{EfficientNetB7} &
  $299\times299$ &
  $3079\times3079$ &
  &
  \textbf{ResNet152\textsuperscript{$\dagger$} } &
  $96\times96$ &
  $1451\times1451$ \\ \cline{1-3} \cline{5-7}

  \textbf{VGG11} &
  \multicolumn{2}{c|}{$150\times150$} &
  &

  \textbf{GoogLeNet} &
  $123\times123$ &
  $379\times379$ \\ \cline{1-3} \cline{5-7}


  \textbf{VGG13} &
  \multicolumn{2}{c|}{$156\times156$} &
  &

  \textbf{SqueezeNet 1.0} &
  $67\times67$ &
  $155\times155$ \\ \cline{1-3} \cline{5-7}

  \textbf{VGG16} &
  \multicolumn{2}{c|}{$212\times212$} &
  &
  \textbf{SqueezeNet 1.1} &
  $63\times63$ &
  $207\times207$ \\ \cline{1-3} \cline{5-7}

  \textbf{VGG19} &
  \multicolumn{2}{c|}{$268\times268$} &
  &
  \textbf{InceptionResNetV2} &
  $143\times143$ &
  $3039\times3039$ \\ \cline{1-3} \cline{5-7}

  \textbf{Xception} &
  $135\times135$ &
  $1083\times1083$ &
  &
  \textbf{InceptionV3} &
  $239\times239$ &
  $1311\times1311$ \\ \cline{1-3} \cline{5-7}
\end{tabular}}
\end{table*}
\includegraphics[]{pngSeed}

\end{document}